\documentclass[final,3p,times,twocolumn]{elsarticle}

\usepackage{graphicx}
\usepackage{amssymb}

\usepackage{lineno}

\journal{Nuclear Instruments and Methods in Physics Research Section A}

\begin{document}
\begin{frontmatter}

\title{Overview of the CLEAR plasma lens experiment}

\author[UniOslo]{C. A. Lindstr{\o}m}
\ead{c.a.lindstrom@fys.uio.no}
\author[UniOslo]{K. N. Sjobak}
\author[UniOslo]{E. Adli}
\address[UniOslo]{Department of Physics, University of Oslo, 0316 Oslo, Norway}

\author[DESY]{J.-H. R{\"o}ckemann}
\author[DESY]{L. Schaper}
\author[DESY]{J. Osterhoff}
\address[DESY]{DESY, Notkestra{\ss}e 85, 22607 Hamburg, Germany}

\author[UniOxford]{A. E. Dyson}
\author[UniOxford]{S. M. Hooker}
\address[UniOxford]{University of Oxford, Clarendon Laboratory, Parks Road, Oxford OX1 3PU, United Kingdom}

\author[CERN]{W. Farabolini}
\author[CERN]{D. Gamba}
\author[CERN]{R. Corsini}
\address[CERN]{CERN, Geneva, Switzerland}

\begin{abstract}
Discharge capillary-based active plasma lenses are a promising new technology for strongly focusing charged particle beams, especially when combined with novel high gradient acceleration methods. Still, many questions remain concerning such lenses, including their transverse field uniformity, limitations due to plasma wakefields and whether they can be combined in multi-lens lattices in a way to cancel chromaticity. These questions will be addressed in a new plasma lens experiment at the CLEAR User Facility at CERN. All the subsystems have been constructed, tested and integrated into the CLEAR beam line, and are ready for experiments starting late 2017.
\end{abstract}

\begin{keyword}
Active plasma lens \sep Discharge capillary \sep Compact Marx Bank \sep CLEAR User Facility
\end{keyword}

\end{frontmatter}


\section{Introduction}
Recent novel accelerator research has delivered several intriguing technologies, some of which can provide accelerating fields of several GV/m, paving the way to building significantly more compact particle accelerators. Many of these emerging concepts, e.g. plasma wakefield accelerators \cite{ChenPRL1985,RuthPA1985} or direct laser accelerators \cite{PeraltaNature2013}, require tightly focused input beams. Unless the components used to focus particle beams are also made more compact, not much will be gained from these advances.

Active plasma lensing is a promising technique providing strong and compact focusing, and has already been used in high gradient accelerator staging \cite{SteinkeNature2016}. After breaking down a diffuse gas to a plasma, a large on-axis current density is used to form a radially increasing azimuthal magnetic field which, unlike a quadrupole, provides focusing in both planes simultaneously. Such a device, typically a thin gas filled capillary with high voltage electrodes on either side, can readily create magnetic field gradients upwards of kT/m \cite{TilborgPRL2015}.

Although finding stronger focusing elements is necessary, it is not sufficient for compact staging of advanced accelerator structures. Highly divergent beams and \%-level energy spreads imply that the chromaticity of the staging optics needs to be controlled \cite{LindstromNIMA2016}. Typically this requires sextupoles, however these magnets have other adverse effects on the beam. The newly developed concept of apochromatic focusing \cite{LindstromPRAB2016} is therefore of interest, where only a lattice of linear optics lenses (like quadrupoles or active plasma lenses) are used to cancel chromaticity at specific locations in the beam line. In this paper, we outline an ongoing experiment aimed at eventually demonstrating such an apochromatic lattice of active plasma lenses, as a path towards compact and chromatically controlled staging of high gradient accelerators.

\section{Experimental goals}
The ultimate goal of demonstrating an apochromatic plasma lens lattice will be approached in two phases: (1) by commissioning and characterizing a single plasma lens and (2) by using three such lenses and measuring their chromaticity as well as alignment and synchronization tolerances. The first of these two phases is underway at the CLEAR Test Facility at CERN, and will investigate two important limitations of active plasma lenses: radially nonuniform focusing fields and interference from plasma wakefields.

\subsection{Successful operation of a novel, low-cost setup}
Several implementations of active plasma lenses have already been successfully demonstrated by LBNL \cite{TilborgPRL2015}, INFN \cite{PompiliAPL2017} and DESY \cite{RockemannTBP2018}. Emphasis is therefore placed on developing a novel low-cost, scalable setup. This is attempted in two ways: using a Marx bank instead of the bulkier and more expensive thyratron, and using a thin polymer foil beam window instead of differential pumping to ensure sufficient vacuum in the rest of the accelerator. Our goal is to demonstrate stable operation over tens of thousands of shots.

\subsection{Field gradient uniformity measurements}
Transporting charged particle beams with a well defined energy without emittance growth requires linear beam optics. Any nonlinearities will, without compensation, lead to emittance growth. Active plasma lenses are ideally linear, but in practice they may have nonlinearities due to e.g.~low partial ionization \cite{PompiliAPL2017} or radial temperature gradients. For the latter case, reference \cite{TilborgPRAB2017} develops a theoretical model for such temperature gradient-based nonlinearities, found consistent with their indirect experimental measurements (halo-formation). It is, however, important to verify the model further by measuring the nonlinearity directly (magnetic field vs.~radius) before attempting to reduce or compensate for it.

In order to directly measure the uniformity of the focusing field, we will observe the centroid angular deflection of a transversely offset beam. This requires a tightly focused beam (compared to the plasma lens aperture) in order to not sample a large range of radii. A short lens should also be used to avoid transverse displacement of the beam inside the lens.

\subsection{Limits set by plasma wakefield focusing}
Plasma wakefields are intrinsically much stronger (MT/m) than those reachable in an active plasma lens (kT/m), but are generally both longitudinally and transversely nonuniform. Recent active plasma lens experiments have largely avoided interference from plasma wakefield focusing (often called passive plasma lensing) by using low charge, large beam sizes or long bunches. However, these low density beams are not representative of the beams planned for high intensity machines like a linear collider, and it is important to understand whether active plasma lenses can be used in such applications. Passive plasma lensing has already been demonstrated experimentally in an active plasma lens \cite{MarocchinoAPL2017}, but this should be further probed to determine what parts of the parameter space allows distortion free active plasma lensing.

We will perform this measurement over a wide range of charge, beam size, bunch length and plasma density, by looking for distortion of the beam in the presence of a plasma, but with no active plasma lens current. This is verified by transversely offsetting the beam in the plasma lens, and looking for focusing without a centroid angular deflection.

\begin{figure}[t]
	\centering\includegraphics[width=0.95\linewidth]{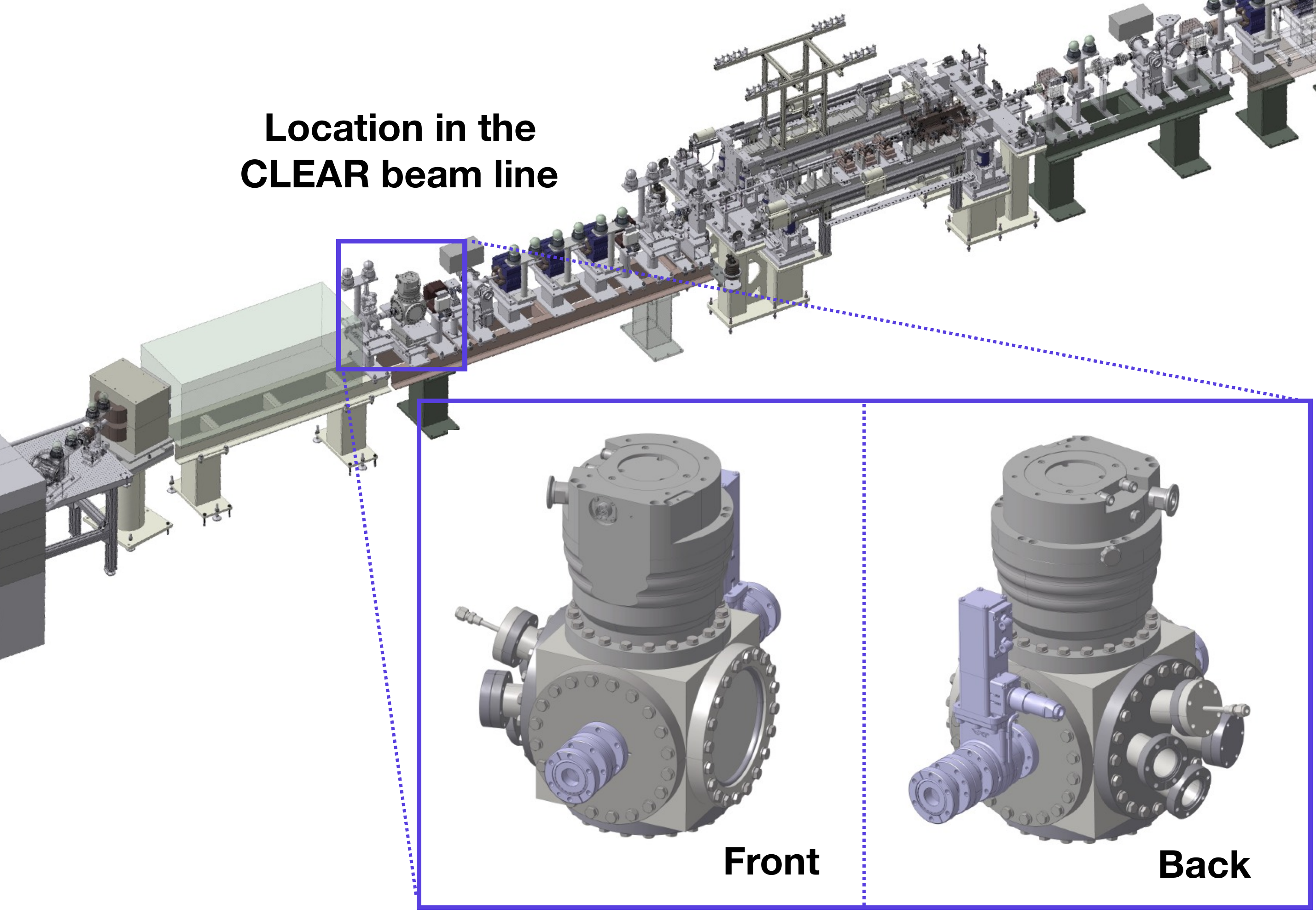}
	\caption{Location of the plasma lens experimental setup as installed in the CLEAR beam line. The inset shows a 3D sketch of the experimental chamber setup.}
    \label{fig:Sketch3D}
\end{figure}

\section{The CLEAR User Facility}
The CERN Linear Electron Accelerator for Research (CLEAR) \cite{GambaNIMA2018} is a user facility well suited for an active plasma lens experiment, due to its versatility and rapid turnaround. Previously used as the probe beam injector for the CLIC Test Facility (CTF3), it uses a photocathode RF gun and three S-band accelerating structures to provide 50-220~MeV electron bunches to a dedicated experimental area. It can produce trains of up to a few hundred bunches with 1~pC to 1.5~nC of charge per bunch, at a repetition rate of maximum 5~Hz. The emittance of these bunches ranges from 3~mm~mrad at 50~pC to 20~mm~mrad at 1~nC. The bunch length can be varied between 300~$\mu$m and 1200~$\mu$m (1-4~ps).

Just upstream of the plasma lens experiment (see Fig.~\ref{fig:Sketch3D}) there is a quadrupole triplet to provide tightly focused beams in the capillary. Calculations indicate beta functions of less than 10~cm such that a minimum rms beam size of 50~$\mu$m should be achievable.

\section{Experimental setup}
\begin{figure}[t]
	\centering\includegraphics[width=0.95\linewidth]{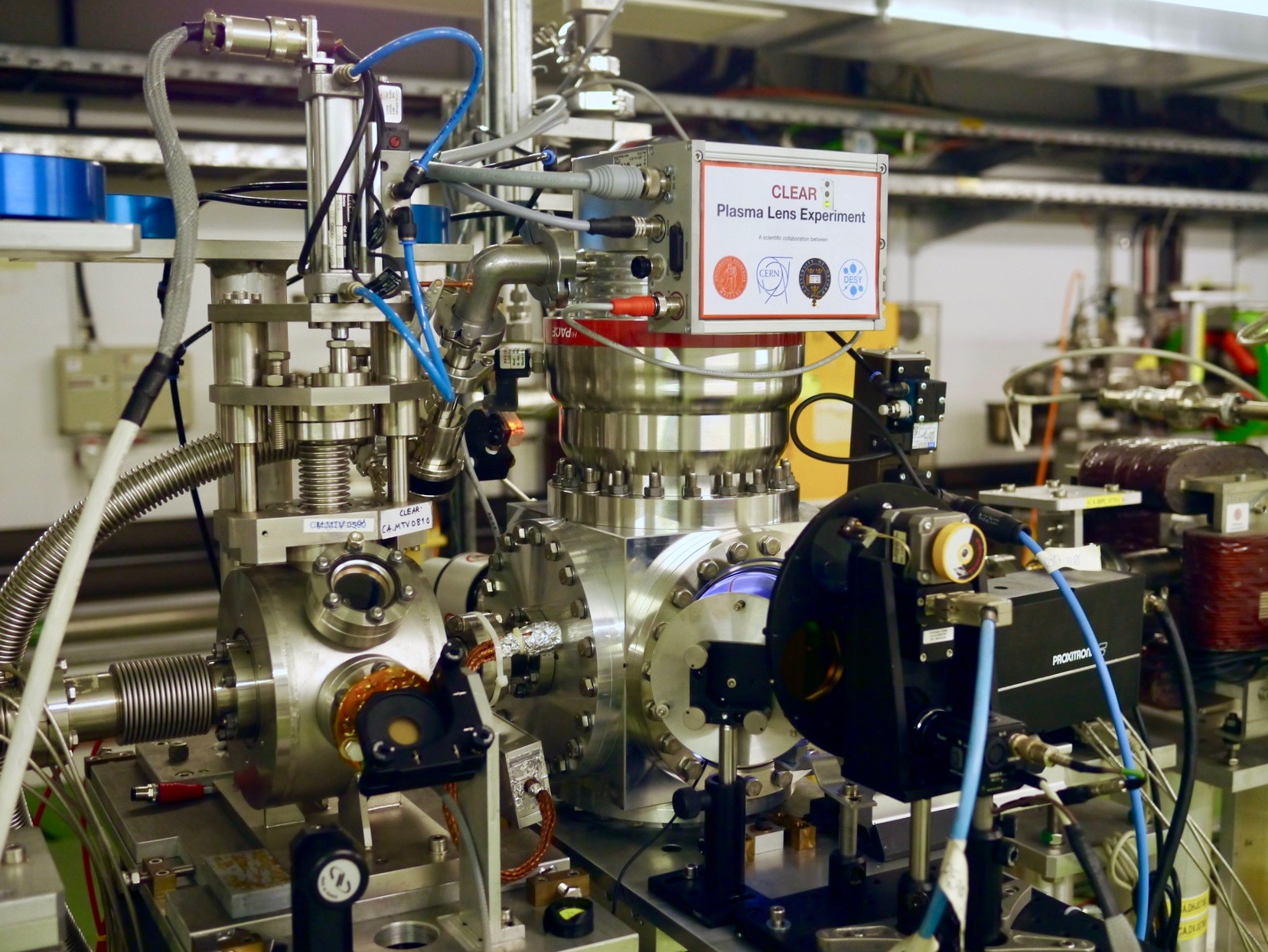}
	\caption{Overview of the installed setup (beam direction: right to left). A cubic vacuum chamber is mounted on a precision mover, and is connected to the beam line via flexible bellows. A large turbo pump is mounted above. A viewport is used to view the capillary mounted inside, captured on both a regular and a 5~ns gated camera. An OTR screen is insertable in a smaller chamber just downstream. A low pressure gas injection system is mounted behind the chamber, as well as a Compact Marx Bank providing high voltage, high current pulses for discharging. A gate valve with a thin polymer foil is used to avoid gas leaking upstream.}
    \label{fig:SetupPicture}
\end{figure}

Although small in size, the experimental setup (Fig.~\ref{fig:SetupPicture}) consists of several subsystems, all of which must work in unison to focus an electron beam.

\begin{figure}[t]
	\centering\includegraphics[width=0.95\linewidth]{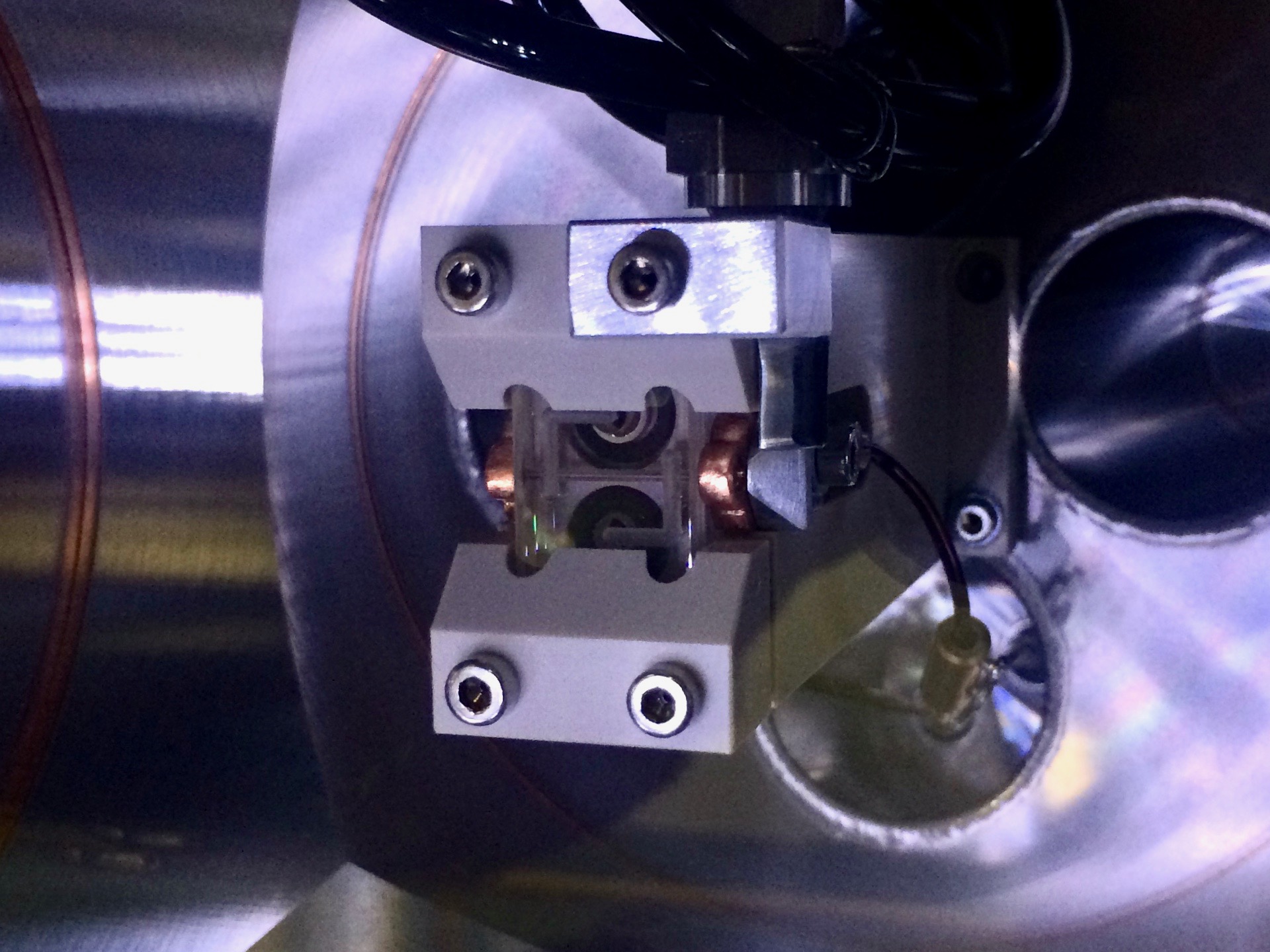}
	\caption{Plasma lens sapphire capillary (1~mm diameter, 15~mm long) mounted in a PEEK holder, with connected copper electrodes. Gas inlet pipes internally in the holder and in the sapphire allow gas to flow from the external low pressure gas injection system into the capillary. Small surfaces angled at 45$^\circ$ on the upstream side produces OTR light for measuring of the beam size locally.}
    \label{fig:PlasmaLensPicture}
\end{figure}

\subsection{Capillary and holder}
The capillary itself is a 1~mm diameter, 15~mm long sapphire tube. A half tube is milled from each of two sapphire blocks using a drill bit (and not using laser ablation), as well as two separate gas inlet lines extending to the edges of the capillary. These gas lines continue into a capillary holder made from polyether ether ketone (PEEK): an ultra-high vacuum compatible, electrically insulating plastic. The internal gas lines in the holder exit via a non-conducting polyurethane gas pipe to a gas feedthrough out of the vacuum chamber inside which the capillary and holder is mounted (see Fig.~\ref{fig:PlasmaLensPicture}). Two electrodes with a hole slightly larger than the capillary diameter are mounted on the upstream and downstream sides of the capillary, and connected to the outside via an electrical feedthrough.

\subsection{Chamber and alignment}
A 20$\times$20$\times$20~cm$^3$ cubic aluminium vacuum chamber is used to electrically insulate the plasma lens from ground with approximately 8 cm of vacuum. The six DN160 CF flanges are used for connecting to the beam line (upstream and downstream), to a turbo pump (above), a borosilicate glass viewport (front), a supporting precision mover (below) and to a multi-port cluster flange (back). The PEEK holder is mounted on the cluster flange, which has four feedthroughs used for two high voltage electric leads rated for 25~kV DC, for gas, and for a pressure gauge measuring the chamber vacuum level.

Alignment and transverse movement is accomplished with a precision two-axis mover with $\mu$m resolution and range 13~mm horizontally and 8~mm vertically. The mover is external to increase movement range, avoid controls interference from the electric discharges and to save on cost. Angular pitch and yaw is manually aligned using adjustment screws on the mover and on the rotating mover-to-chamber connection plate. Longitudinal translation and angular roll adjustment is not necessary. The chamber is connected to the beam pipe with two flexible edge-welded bellows on each side, and the 13~mm horizontal range of the mover is sufficient to move the sapphire capillary completely out of the beam orbit.

\subsection{Vacuum and gas flow}
In this experiment we use both helium and argon, piped via gas lines from 200 bar, 50 l gas bottles outside the accelerator hall. These lines connect to a low-pressure precision gas injection system, consisting of a remotely controlled gas flow regulator (Pfeiffer EVR 116) and a buffer volume with a pressure gauge (Pfeiffer CMR 361). This capacitance-based gauge operates in a feedback loop with the gas flow regulator controller (Pfeiffer RVC 300). Gas pressures from 1 mbar up to 1 bar can be kept in this buffer volume, which is connected to the vacuum chamber gas feedthrough via a remotely controllable pneumatic shutter valve. A short gas line ensures that the pressure inside the capillary stays close to the measured pressure in the buffer. The gas is injected continuously to a operate at a stable, known pressure.

\begin{figure}[t]
	\centering\includegraphics[width=0.95\linewidth]{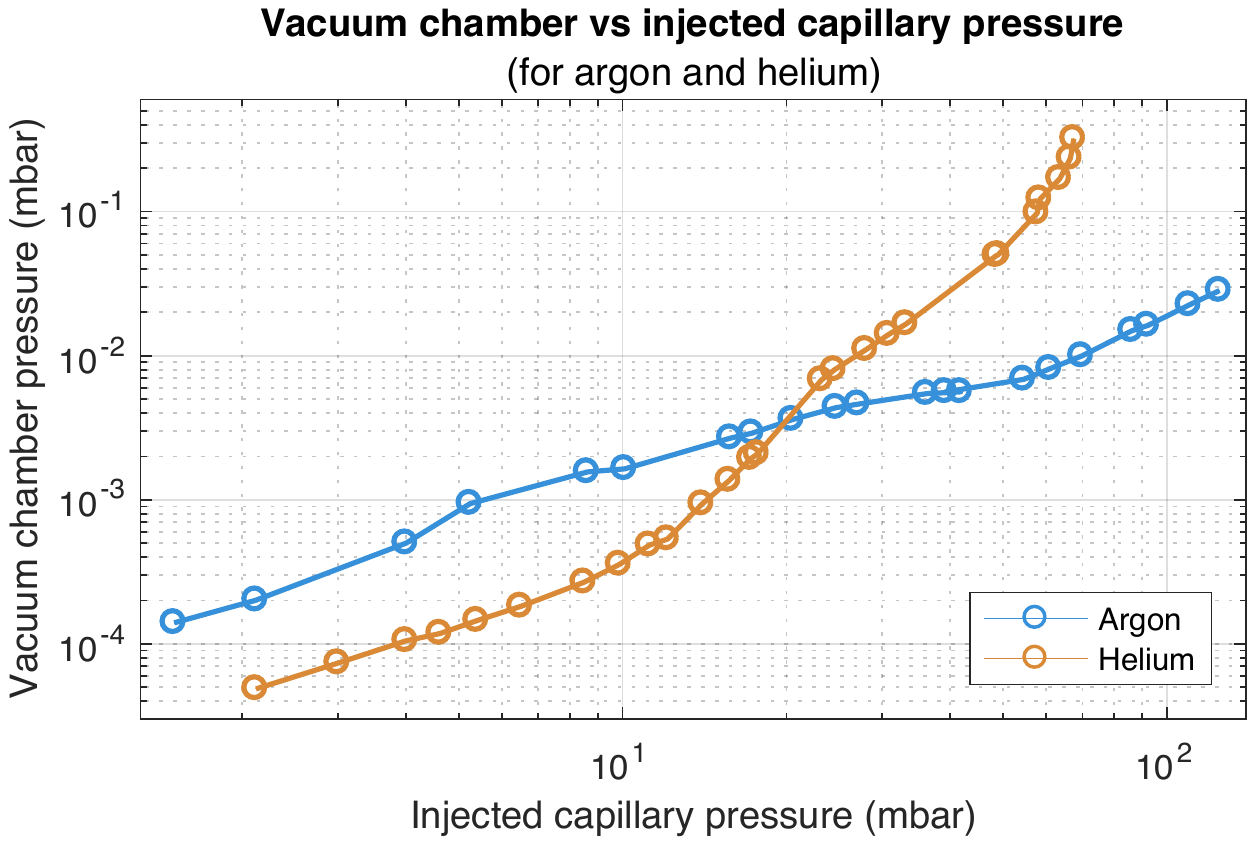}
	\caption{Measurement of vacuum chamber pressure vs injected capillary pressure, for both helium and argon. Avoiding sparks to the chamber wall requires chamber pressures below 0.01~mbar, indicating safe operation up to about 30~mbar in the capillary for helium, and 70~mbar for argon.}
    \label{fig:GasLevels}
\end{figure}

Inside the chamber, the gas which escapes the small aperture of the capillary must be rapidly pumped out. It is important to keep the chamber pressure below approximately 0.01~mbar to avoid discharging to ground rather than between the electrodes. A large turbo pump with magnetic bearings (Pfeiffer HiPace 700M) and a pumping speed of 700~l/s ensures a sufficient vacuum for capillary pressures up to 30~mbar of helium or up to 70~mbar of argon (see Fig.~\ref{fig:GasLevels}). The turbo pump is then connected in series with a 15~m$^3$/h scroll pump (Edwards nXDS15i) to ensure a fore vacuum of less than 0.1~mbar.

The chamber pressure is measured using a full range Pirani/cold cathode gauge (Pfeiffer PKR 361), which allows measurement down to 10$^{-9}$~mbar. Typically the turbo pump reaches 10$^{-8}$~mbar with no gas flow.

\subsection{Polymer foil beam window}
During operation of the CLEAR accelerator, the photocathode requires a very good vacuum and any gas flow upstream is unacceptable. Typically this is solved by differential pumping, which is expensive (many pumps) and requires much space. Instead, based on experience at the PITZ experiment at DESY Zeuthen \cite{LishilinNIMA2016}, we have installed a thin polymer beam window just upstream (20~cm) of the plasma lens. Made from an 8~$\mu$m thick Kapton foil mounted in a small retractable gate valve, it can withstand pressure differentials of up to 1 bar. Early tests shows negligible gas permeation, but that the beam experiences a slight increase in beam emittance when passing through the window. Unless the chamber is erroneously filled with 1~bar of gas, the window never experiences large forces. The plan is therefore to change to a thinner 3~$\mu$m Kapton or Mylar foil, which reduces the scattering of the beam.

\subsection{Compact Marx Bank high voltage discharge source}
\begin{figure}[t]
	\centering\includegraphics[width=0.9\linewidth]{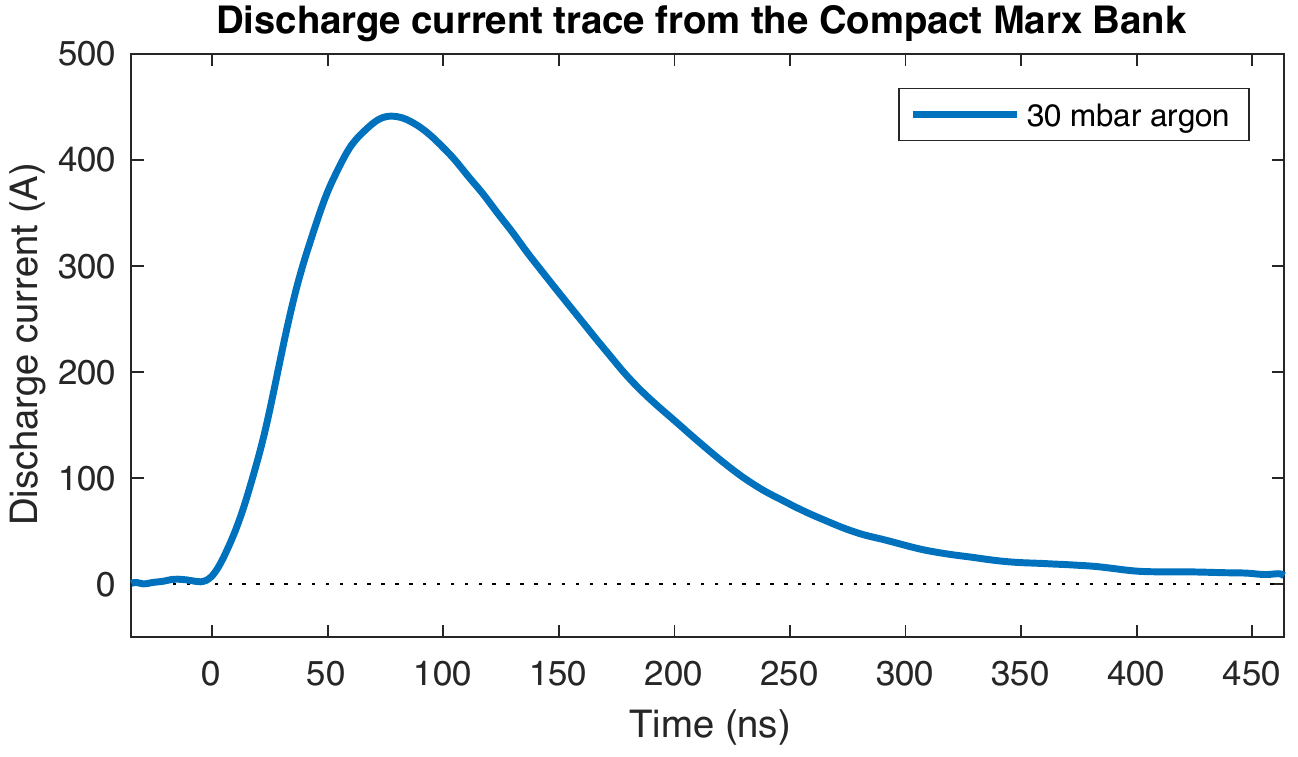}
	\caption{Current trace (300 shots average) from the Compact Marx Bank during a discharge in the capillary, measured using a wide-band current pulse transformer, showing a peak current of 450~A. The gas pressure in the capillary was 30~mbar of argon.}
    \label{fig:MarxBankCurrentTrace}
\end{figure}

To break down the gas to a plasma and to supply the large current required for beam focusing, a 10-stage spark-gap based Compact Marx Bank \cite{DysonRevSci2016} is used. When triggered, it releases a 20~kV sub-$\mu$s duration pulse of peak current up to 500~A. Two wide-band current transformers (Pearson 410) are used to measure both the incoming and outgoing current pulses (see Fig.~\ref{fig:MarxBankCurrentTrace} for a representable current trace). It is important to note that such a high current, high voltage source poses a significant safety risk, and must be handled accordingly.

\subsection{Timing and synchronization}
The discharge from the Compact Marx Bank must be synchronized with the passing of the beam to within a few ns. In addition, the relative timing of these two events should cover a range of $\mu$s in order to scan the beam through the full current pulse profile. This is accomplished by connecting the discharge first to a coarse trigger with 52~ns step resolution and a large range, and then through a smaller range fine delay trigger with 4~ps step resolution.

\subsection{Diagnostics}
Several diagnostics are used to measure the output of the experiment. To detect changes in beam size and any dipole kicks in the plasma lens is an insertable optical transition radiation (OTR) screen 30~cm downstream, with an image resolution of approximately 20~$\mu$m/pixel, and a thin aluminum blinder foil just upstream to stop any plasma light. Another camera looks directly at the sapphire capillary through the viewport, showing the transverse and longitudinal profile of the discharge in the capillary as well as any scintillation light from the beam passing through sapphire. A gated camera with a minimum gate duration of 5~ns is used to observe the temporal variation of the discharge, needed to understand the evolution of the plasma. Immediately downstream of the experimental chamber a photomultiplier tube (PMT) is installed to observe particle losses in the capillary. Further downstream is a charged particle beam dipole spectrometer to measure any energy changes caused by plasma wakefields, although this is expected to be a negligible effect. An in-air yttrium aluminum garnet (YAG) screen is used for measuring the transverse profile several meters downstream, before the beam is dumped. Lastly, two small OTR-producing surfaces are mounted on the upstream electrode to provide a beam size measurement as close to the capillary as possible, important for minimizing the beam size.

\begin{figure}[t]
	\centering\includegraphics[width=0.95\linewidth]{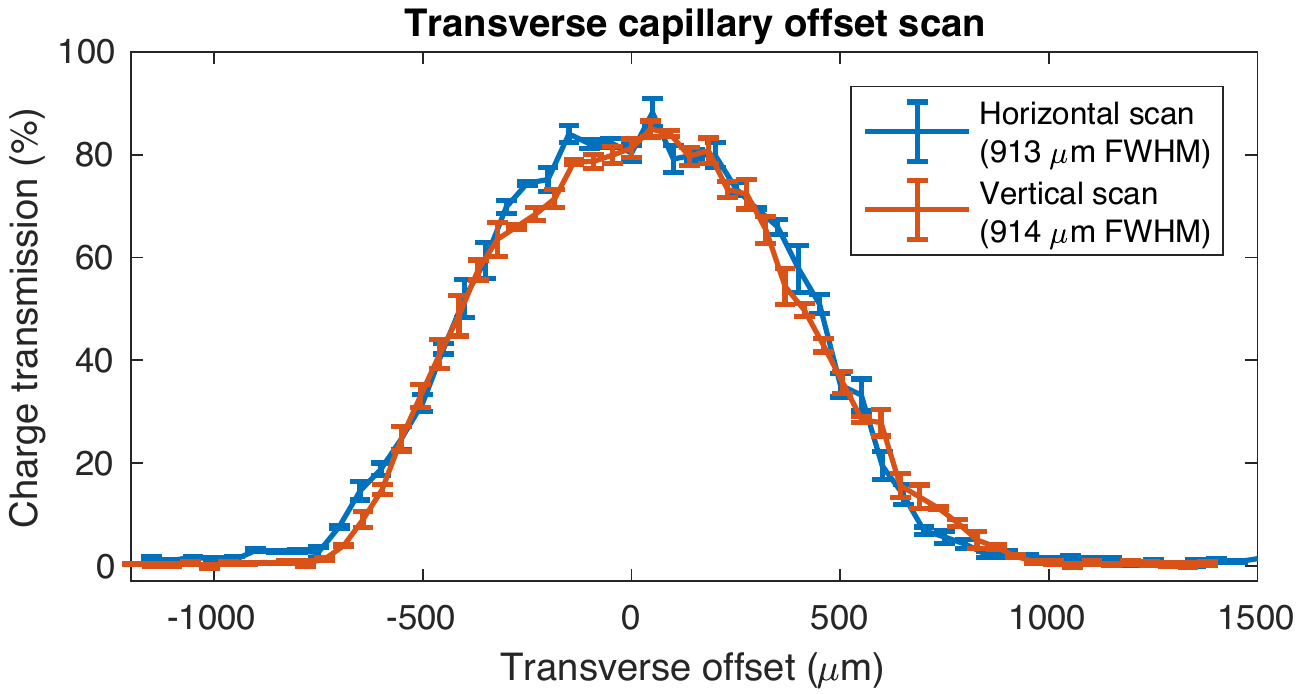}
	\caption{Transmission scan of the beam through the 1~mm diameter capillary, both in the horizontal (blue) and the vertical direction (red), where the charge transmission is calculated as the ratio to the charge measured with the plasma lens completely out. The full width at half maximum (FWHM) corresponds to the capillary radius, but is slightly less due to small angular misalignment. Simulations indicate a beam size of 170$\pm$10~$\mu$m rms in both planes, which can be decreased to less than 50~$\mu$m with further optimization.}
    \label{fig:TransmissionScan}
\end{figure}

\section{Status and future plans}
The CLEAR plasma lens experiment started its design phase in early 2017 and is planned to last until at least late 2018.

Bench tests of the vacuum levels, the polymer beam window, gas injection, and high voltage discharges were all successfully performed during mid 2017. The setup was subsequently installed in the CLEAR beam line and integrated into the control system. Some distortion of the beam due to the plasma-current was observed during a preliminary beam time in late 2017, before further upgrades were initiated. The best charge transmission of the beam through the capillary is currently $87\pm5$\% (see Fig.~\ref{fig:TransmissionScan}), but this number is expected to reach close to 100\% as the beam size approaches the stipulated 50~$\mu$m rms. Experiments will be performed starting at the end of 2017 and is planned to last until mid 2018.

\section{Conclusion}
The CLEAR plasma lens experiment is a new experiment aimed at demonstrating successful operation of a novel low-cost setup, and at characterizing two important aspects of active plasma lenses: the non-uniformity of the focusing field, and limits due to plasma wakefields. After successful bench tests of all the subsystems, the setup is now installed and starting to produce scientific data. Experiments will continue until late 2018.

\section{Acknowledgements}
The authors wish to thank Reidar Lunde Lillestøl, Gianfranco Ravida, Gerard McMonogle, Franck Perret, Stephane Curt, Thibaut Lefevre, Bruno Cassany, Serge Lebet, Jose Antonio Ferreira Somoza, Alice Ingrid Michet, Herve Rambeau, Stefano Mazzoni, Michael John Barnes, Aimee Ross and Candy Capelli. We thank CERN for providing beam time at the CLEAR User Facility. This work was supported by the Research Council of Norway (NFR Grant No. 230450) and by the Helmholtz Association of German Research centers (Grant No. VH-VI-503).


\end{document}